\documentclass{article}

\usepackage{amsmath,amsthm,xcolor,hyperref,graphicx,amssymb}
\usepackage{authblk}

\theoremstyle{thmstyleone}%

\theoremstyle{thmstyletwo}%

\theoremstyle{thmstylethree}%

\usepackage{listings}

\definecolor{myviolet}{rgb}{0.6,0.0,0.65}
\definecolor{myblue}{rgb}{0.1,0.0,0.8}
\definecolor{mygreen}{rgb}{0,0.4,0.2}
\definecolor{mycomment}{rgb}{0.7,0.2,0.2}
\definecolor{myred}{rgb}{0.8,0.0,0.0}
\definecolor{myorange}{rgb}{0.6,0.2,0.2}

\lstdefinelanguage{coq}[]{Caml}{
keywords=[1]{Section,Definition,Defined,CoInductive,Coercion,Inductive,Variant,Fixpoint,
  Parameter,Local,Module,Import,Record,Structure,Axiom,Lemma,Corollary,Proposition,Theorem,Notation,
  Reserved,End,Proof,Goal,Qed,Variable,Variables,Context,Implicit,Types,Hypothesis,Hypotheses,Let,Program,Canonical,Check,Fail,Example,Declare,Scope,Delimit},
keywordstyle=\color{myviolet}\ttfamily,
morekeywords=[2]{match,with,end,Set,Prop,Type,fun,of,let,in,struct,if,is,then,else,return},
keywordstyle=[2]\color{mygreen}\ttfamily,
morekeywords=[3]{move,refine},
keywordstyle=[3]\color{myblue}\ttfamily,
morekeywords=[3]{elim,field,lra,rewrite,HB,mixin,structure,instance,factory,builders},
keywordstyle=[4]\color{myred}\ttfamily,
morekeywords=[4]{by,reflexivity,exact}
}

\def\pipe{\char`\|}

\let\L=\lstinline

\begin{document}

\def\mathcomp{\textsc{MathComp}}
\def\ssreflect{\textsc{SSReflect}}
\def\coq{\textsc{Coq}}
\def\lean{\textsc{Lean}}
\def\hollight{\textsc{HOL-Light}}
\def\agda{\textsc{Agda}}
\def\nuprl{\textsc{Nuprl}}
\def\metamath{\textsc{Metamath}}
\def\mathlib{\textsc{mathlib}}
\def\analysis{\textsc{MathComp-Analysis}}
\def\hb{\textsc{Hierarchy-Builder}}
\def\finmap{\textsc{Finmap}}
\def\coquelicot{\textsc{Coquelicot}}
\def\companycoq{\textsc{Company-Coq}}

\def\us{\char`\_}

\def\salgebra{$\sigma$-\textsl{algebra}}
\def\sadditive{$\sigma$-\textsl{additive}}
\def\ssubadditive{$\sigma$-\textsl{subadditive}}
\def\sadditivity{$\sigma$-\textsl{additivity}}
\def\ssubadditivity{$\sigma$-\textsl{subadditivity}}
\def\sfinite{$\sigma$-\textsl{finite}}

\def\rei#1{{\color{red}{[NB(rei): #1]}}}
\def\cyr#1{{\color{blue}{[NB(cyr): #1]}}}

\def\newterm#1{\textsl{#1}}
\def\sect#1{Sect.\,#1}
\def\measurable#1{\texttt{measurable}(#1)}
\def\mydef{\overset{\textrm{def}}{=}}

\title{Measure Construction by Extension in Dependent Type Theory with Application to Integration}
\date{}

\author[1]{Reynald Affeldt}
\author[2]{Cyril Cohen}
\affil[1]{National Institute of Advanced Industrial Science and Technology (AIST)}
\affil[2]{Universit\'e C\^ote d'Azur and Inria}

\maketitle

\begin{abstract}

We report on an original formalization of measure and integration
theory in the Coq proof assistant.
We build the Lebesgue measure following a standard construction that
had not yet been formalized in proof assistants based on dependent type theory: by
extension of a measure over a semiring of sets.
We achieve this formalization by leveraging on existing techniques
from the Mathematical Components project.
We explain how we extend Mathematical Components' iterated operators
and mathematical structures for analysis to provide support for
infinite sums and extended real numbers.
We introduce new mathematical structures for measure theory and
incidentally provide an illustrative, concrete application of
Hierarchy-Builder, a generic tool for the formalization of hierarchies
of mathematical structures.
This formalization of measure theory provides the basis for a new
formalization of the Lebesgue integration compatible with
the Mathematical Components project.

\end{abstract}

\section{Introduction}
\label{sec1}

Measure theory and integration theory are major topics in mathematics
with practical applications.  For example, they serve as the
foundation of probability theory whose formalization in proof
assistants is used to verify information security (e.g.,
\cite{abate2021csf}) or artificial intelligence (e.g.,
\cite{tassarotti2021cpp}). It is therefore no wonder that the topic of
formalization of measure and integration theory in proof assistants
has already been tackled several times (e.g.,
\cite{bialas1992jfm,bialas1995jfm,hurd2002phd,coble2010phd,holzl2011itp}).
In fact, experiments are still going on~\cite{vanDoorn2021itp}, some still dealing with the
basics~\cite{endou2020jfm,boldo2021jar}.

Our motivation is to develop measure and integration theory on top of
\mathcomp{}~\cite{mathcomp}, a library of formalized mathematics
developed using the \coq{} proof assistant~\cite{coq}.
The \mathcomp{} library consists of several algebraic theories that made it
possible to formalize the Odd Order theorem\footnote{The Odd Order theorem,
a.k.a.\ the Feit-Thompson Theorem, states that groups of odd order are solvable.
This theorem relies on finite group theory, character theory and galois
theory.} by following its published, revised
proof~\cite[\sect{6}]{gonthier2013itp}.
There are now several libraries that are built on top of \mathcomp{},
the main ones being made available as parts of the Mathematical Components
project\footnote{\url{https://github.com/math-comp/}}.
Among them, \analysis{}~\cite{cohen2018jfr,affeldt2020ijcar} aims at
taking advantage of the algebraic theories provided by \mathcomp{}
to develop classical analysis (topology, real and complex analysis,
etc.).

In this paper, we report on an original formalization of measure and
integration theory.
Our approach is to extend \analysis{} with reusable theories while
following textbook presentations~\cite{klenke2014,liintegration}.
The best illustration is the construction of the Lebesgue measure that
we formalize. This is a standard construction from a
semiring of sets, using the Measure Extension theorem.
To the best of our knowledge, it has never been formalized
with the abstraction of ring of sets or semiring of sets
in a proof assistant based on dependent type theory.
Yet, its formalization is the occasion to develop new
mathematical structures of general interest for \coq{} users.
Similarly, the construction of the Lebesgue integral gives us the
opportunity to develop a generic formalization of simple functions and
to extend the formalization of the iterated operators of
\mathcomp{}~\cite{bertot2008tphols}, one key to the successful
formalization of the Odd Order theorem.

Our contribution in this paper is twofold.
First, we bring to the \coq{} proof assistant a formalization of
measure and integration theory that is compatible with the algebraic
theories of \mathcomp{}.
Second, we demonstrate recent formalization techniques developed in
the context of the Mathematical Components project.
In particular, we use \hb{}~\cite{cohen2020fscd} to formalize a
hierarchy of mathematical structures for measure theory and to provide
a compositional formalization of simple functions.
Our technical contributions materialize as extensions to \analysis{}
in the form of reusable formal theories about sequences (of reals and
of extended real numbers) and about sums over general sets and over
finitely-supported functions.

\paragraph*{Paper Outline}
In Sect.~\ref{sec:ereal}, we explain how we develop the theory of
extended real numbers by extending \mathcomp{} and \analysis{}.
In Sect.~\ref{sec:measure_theory}, we explain how we encode the basic
definitions of measure theory, demonstrating the use of
\hb{}.
In Sect.~\ref{sec:extension}, we formalize the Measure Extension
theorem which shows how to extend a measure over a semiring of sets to
a \salgebra{}. This is a standard and generic approach to the
construction of measures.
In Sect.~\ref{sec:lebesgue_measure}, we obtain the Lebesgue measure
by extending a measure over the semiring of sets of intervals.
In Sect.~\ref{sec:lebesgue_integral}, we show that the framework
developed so far allows for a formalization of the Lebesgue integral
up to the dominated convergence and Fubini's theorems.
We review related work in Sect.~\ref{sec:related_work} and conclude
in Sect.~\ref{sec:conclusion}.

\def\accompanying#1{\cite[file \L!#1!]{analysis}}

\paragraph{Note on Notation}
The Mathematical Components project have been favoring ASCII notations.
Most of them are unsurprising because they are inspired by \LaTeX\ commands.
This paper follows this tradition;
ASCII notations will be explained in the prose
or in Tables~\ref{tab:iterated} and~\ref{tab:set}
As a consequence, we can display the \coq{} code almost verbatim;
we allow pretty-printing only for a few standard symbols (such as
\L!<-! instead of {\tt <-},
\L!->! instead of {\tt ->},
\L!forall! instead of {\tt forall},
\L!exists! instead of {\tt exists},
\L!<=! instead of {\tt <=},
\L+!=+ instead of {\tt !=},
\L!/\! instead of \verb!/\!,
etc.).
The accompanying development~\cite{analysis} can be found online and
we will refer to it as a citation possibly indicating the name of the
relevant file (as in \accompanying{filename.v}).

\section{Support for Extended Real Numbers}
\label{sec:ereal}

Since a measure is potentially infinite, it is represented by extended
real numbers. A prerequisite for the construction of measures is
therefore the development of the theory of extended real numbers and
of their sequences. This actually calls for a substantial extension of
\analysis{}~\cite{cohen2018jfr}.

\begin{figure}[htbp]
\centering
\includegraphics[width=2.5cm]{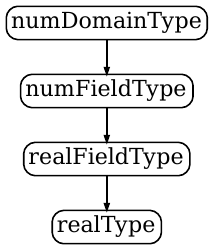}
\caption{Numeric types provided by \mathcomp{} and \analysis{} used in this paper}
\label{fig:numtypes}
\end{figure}

Our starting point is the hierarchy of numeric and real interfaces
provided by \mathcomp{} and \analysis{}.  It contains (among others) the
type \L!numDomainType!  for numeric integral domains, the type
\L!numFieldType! for numeric fields, the type \L!realFieldType!  for
real fields (see~\cite[Chapter~4]{cohen2012phd}), and the type
\L!realType! for real numbers). They form an inheritance chain as
depicted in Fig.~\ref{fig:numtypes}.

The definition of extended real numbers is unsurprising (and predates
the work presented in this paper):
\begin{lstlisting}
Inductive extended (R : Type) := EFin of R | EPInf | ENInf.
\end{lstlisting}
Hereafter, the notation \L!+oo! (resp.\ \L!-oo!) is for the constructor
\L!EPInf!  (resp.\ \L!ENInf!). The constructor \L!EFin! injects a real
number \L!r! into the set of extended real numbers; we also use the
notation \L!r

\subsection{Algebraic Aspects of Extended Real Numbers}
\label{sec:ereal_algebraic}

The expression $\infty-\infty$ is undefined in the
mathematical practice. How to deal with this is a crucial
aspect of our formalization. We define it to be $-\infty$
because it makes the extended real numbers a commutative monoid,
so we can use \mathcomp{}'s iterated operators~\cite{bertot2008tphols}.

Furthermore, we can combine the iterated operators of \mathcomp{}
the notion of limit, which comes from \analysis{}~\cite{cohen2018jfr},
to introduce a notation for infinite sums.
On the one hand, \mathcomp{} comes with
a generic definition of iterated operators \L!\big[op/idx]_(i <- s | P i) f i!
where \L!f! is a function whose domain corresponds
to the list of indices \L!s! and \L!P! is a boolean predicate.
Depending on the properties of the binary operator \L!op! and the
element \L!idx!, many lemmas are available that have been
key to important formalizations in \coq{} (e.g., \cite{gonthier2013itp}).
The notation \L!\big[op/idx]_(i < n | P i) f i! is a special
case where the indices are the natural numbers less than~\L!n!.
As for the notation \L!\sum_(i <- s | P i) f i!, it is a special
case for the iterated addition when \L!f! is a numerical type-valued function.
On the other hand, \analysis{} comes with a definition of
limit~\cite[\sect{2.3}]{cohen2018jfr}. It can be applied to
sequences, i.e., functions of type \L!nat -> T! (notation \L!T^nat!).
Given a sequence~\L!u!, \L!lim u! is the limit of the sequence $\texttt{u}_n$
when $n \to \infty$.
We combine these two notations into a family of
notations \L!\sum_(i <oo | P i) f i!, which is simply defined as
\L!lim (fun n => \big[op/idx]_(i < n | P i) f i)!.
Of course, these new notations need to be instrumented with many
lemmas, the rest of this paper will provide several examples.
Table~\ref{tab:iterated} contains a summary of the notations for
iterated operators we have discussed so far\footnote{
  Table~\ref{tab:iterated} also contains notations that we will
  introduce later in this paper. We summarize these notations together
  to highlight their resemblances and serve as a reading guide.}.

\makeatletter
\renewcommand{\boxed}[1]{\text{\fboxsep=.2em\fbox{\m@th$\displaystyle#1$}}}
\makeatother

\def\bop{\boxed{\mathtt{op}}}

\begin{table}[htbp]
{\centering
\caption{Summary of iterated operators and alike used or newly introduced in this paper.
The symbol $\bop$ is the iterated operator corresponding to \texttt{op}.
}
\label{tab:iterated}
\begin{center}
\begin{tabular}{|ll|}
\hline
\multicolumn{2}{|l|}{Finitely iterated operators \cite{bertot2008tphols}:} \\
\L!\big[op/idx]_(i <- s | P i) f i! & $\bop_{i < \mid s\mid, i \in P} f(s_i)$ \\
\L!\big[op/idx]_(i < n | P i) f i!  & $\bop_{0 \leq i < n, i \in P} f(i)$ \\
\L!\big[op/idx]_(m <= i < n | P i) f i!  & $\bop_{m \leq i < n, i \in P} f(i)$ \\
\multicolumn{2}{|l|}{Application to numeric functions (see Table~\ref{tab:set} for application to sets):} \\
\L!\sum_(i <- s | P i) f i! & $\sum_{i < \mid s\mid, i \in P} f(s_i)$ \\[0.5ex]
\hline
\multicolumn{2}{|l|}{Countably iterated sum of numeric functions (Sect.~\ref{sec:ereal_algebraic}):} \\
\L!\sum_(i <oo | P i) f i! & $\sum_{i = 0, i \in P}^\infty f(i)$ \\
\L!\sum_(m <= i <oo | P i) f i! & $\sum_{i = m, i \in P}^\infty f(i)$ \\[0.5ex]
\hline
\multicolumn{2}{|l|}{Iterated operators over finite supports (Sect.~\ref{sec:fsbigop}):} \\
\L!\big[op/idx]_(i \in D) f i! & $\bop_{i \in D}f(i)$ if $f(i)$ has a finite number\\
& of values in $D$ s.t.\ $f(i) \neq \mathtt{idx}$ o.w. $\mathtt{idx}$\\
\hline
\multicolumn{2}{|l|}{Sum of extended real numbers over general sets (Sect.~\ref{sec:esum}):} \\
\L!\esum_(i in P) f i! & $\sum_{i \in P} f(i)$ \\[0.5ex]
\hline
\multicolumn{2}{|l|}{Integral (Sect.~\ref{sec:integral_measurable_function}):} \\
\L!\int[mu]_(x in D) f x! & $\int_{x \in D} f(x) d\mu(x)$ \\[0.5ex]
\hline
\end{tabular}
\end{center}
}
\end{table}

\subsection{Topological Aspects of Extended Real Numbers}
\label{sec:ereal_topology}

\analysis{} provides several mathematical structures (topological,
uniform, pseudometric spaces, etc.) together with generic lemmas.
To enjoy these lemmas, it is necessary to equip extended real numbers
with these structures by showing they meet their interfaces.

Extended real numbers form a pseudometric space. The instantiation of
the mathematical structures essentially relies on the
definition and properties of an order-preserving bijective function from
the set of extended real numbers to $[-1;1]$ (see \accompanying{constructive_ereal.v} for details):
\begin{lstlisting}
Definition contract (x : \bar R) : R :=
  match x with r%:E => r / (1 + `|r|) | +oo => 1 | -oo => -1 end.
\end{lstlisting}
There is no hope to get a richer structure (say, \mathcomp's \L!zmodType!) on the full type though,
because as we already discussed above $\infty-\infty$ is taken
to be $-\infty$.

\subsection{Sequences of Extended Real Numbers}
\label{sec:ereal_seq}

The preparatory steps (Sections~\ref{sec:ereal_algebraic} and~\ref{sec:ereal_topology}) we briefly overviewed above are necessary to
produce a theory about sequences of extended real numbers that blends
in \analysis{} in a satisfactory way. For the sake of illustration,
let us present two sample lemmas. The first one shows that the
limit of a sum is the sum of limits:
\begin{lstlisting}
Lemma ereal_limD (R : realType) (f g : (\bar R)^nat) :
  cvg f -> cvg g -> lim f +? lim g -> lim (f \+ g) = lim f + lim g.
\end{lstlisting}
We already explained the notation \L!lim! in Sect.~\ref{sec:ereal_algebraic}.
See Fig.~\ref{fig:numtypes} for \L!realType!.
The definition \L!cvg f! (\L!cvg! is for ``convergence'') states that \L!lim f! exists without naming it
explicitly.
The notation \L!a +? b! is a predicate that checks
whether the addition of \L!a! and \L!b! is well-defined;
the notation \L!f \+ g! is for the pointwise addition of two functions.

The second illustrative lemma shows the commutation of
finite and infinite sums of sequences of non-negative terms:
\begin{lstlisting}
Lemma nneseries_sum_nat (R : realType) n (f : nat -> nat -> \bar R) :
  (forall i j, 0 <= f i j) ->
  \sum_(j <oo) (\sum_(0 <= i < n) f i j) =
  \sum_(0 <= i < n) (\sum_(j <oo) (f i j)).
\end{lstlisting}
There are many lemmas dealing with sequences of extended real numbers
that have been added to \analysis{} for the purpose of this work (see
\accompanying{sequences.v} and \accompanying{normedtype.v}).  These
are reusable lemmas that make the rest of our formalization possible.

\subsection{Iterated Operators over Finite Supports}
\label{sec:fsbigop}

\def\cplt#1{{#1}^{\complement}}

\begin{table}[tbp]
{
\caption{Summary of the set-theoretic notations used in this paper.
The type \L!set T! is defined as \L!T -> Prop!.
Most set-theoretic constructs are given ASCII notations,
otherwise we use the \coq{} identifier directly (as with {\tt set0} or \L!trivIset!).
}
\begin{center}
\begin{tabular}{|l|l|l|l|}
\hline
ASCII                & \coq{}       & Meaning                                      \\
notation             & identifier &                                              \\
\hline
{\tt set0}           & {\tt set0} & The empty set                                \\
\L![set: A]!         & \L!setT!     & The full set of elements of type \L!A!       \\
{\tt `\pipe`}        & \L!setU!     & Set union                                    \\
{\tt `\&`}           & \L!setI! & Set intersection                  \\
\L!`\`!              & \L!setD!     & Set difference                               \\
\L!~`!               & \L!setC!     & Set complement                               \\
{\tt `<=`}           & \L!subset!   & Set inclusion                                \\
\L!f @` A!           & \L!image!    & Image by \L!f! of \L!A!                      \\
\L!f @^-1` A!        & \L!preimage! & Preimage by \L!f! of \L!A!                   \\
\L![set x]!          & \L!set1!     & The singleton set $\{x\}$                    \\
\L![set~ x]!         & see \cite{analysis}     & The complement of $\{x\}$                    \\
\L![set E | x in P]! & see \cite{analysis}     & the set of \L!E! with \L!x! ranging in \L!P! \\
\L!range f!          & see \cite{analysis}     & Image by \L!f! of the full set               \\
\L!\big[setU/set0]_! & see Table~\ref{tab:iterated} & $\bigcup_{i < \mid s\mid, i \in P} f(s_i)$ \\
\L! (i <- s | P i) f i! & & \\
\L!\bigcup_(k in P) F k!  & {\tt bigcup}     & Countable union                              \\
\L!\bigcap_(k in P) F k!  & {\tt bigcap}     & Countable intersection                       \\
\L!trivIset D F!     & {\tt trivIset}   & \L!F! is a sequence of pairwise     \\
                     &           &  disjoint sets over the domain \L!D!       \\
\L![set` p]!         & see \cite{analysis} & Set corresponding to the boolean \\
                     &           & predicate \L!p!  \\
\hline
\end{tabular}
\end{center}
\label{tab:set}
}
\end{table}

To be able to succinctly formalize some proofs relying on iterated
operators, we also extend the library of iterated operators of
\analysis{} with \newterm{iterated operators over finite supports}.
They take the form of the notation \L!\big[op/idx]_(i \in A) f i!
for the iterated application of the operator \L!op! to \L!f i!'s where
\L!i! ranges over \L!A! and \L!f! as a finite support.

The definition of the finite support of a function relies on a theory
about the cardinality properties of sets that was also triggered by
the work presented in this paper. From this theory, we use in
particular the function \L!fset_set! (\L!fset_set A! returns a list
when the set \L!A! is indeed finite and the empty list otherwise)
to define the finite support of a function:
\begin{lstlisting}
Definition finite_support {I : choiceType} {T : Type}
    (idx : T) (D : set I) (F : I -> T) : seq I :=
  fset_set (D `&` F @^-1` [set~ idx] : set I).
\end{lstlisting}
The notation for iterated operators over finite supports combines this
definition with \mathcomp's iterated operators:
\begin{lstlisting}
Notation "\big [ op / idx ]_ ( i '\in' D ) F" :=
    (\big[op/idx]_(i <- finite_support idx D (fun i => F)) F) :
  big_scope.
\end{lstlisting}
The integral of simple functions in Sect.~\ref{sec:sintegral} will
provide a concrete use of this new notation.

\subsection{Sums over General Sets}
\label{sec:esum}

Last, we extend \analysis{} with \newterm{sums over general sets},
i.e.:
$$\sum_{i \in S} a_i \mydef \sup \left\{ \sum_{i \in A} a_i \,\text{\textbar}\, A \textrm{ finite subset of } S \right\}.$$
For that purpose, we introduce the definition \L!fsets S! for the finite sets
included in~\L!S!. It is defined using the predicate \L!finite_set! which is
defined in such a way that \L!finite_set A! when there is a natural number~$n$
such that there is bijection between \L!A! and the set $\{ k \;\pipe\; k < n \}$,
i.e., when the set~\L!A! is finite (see \accompanying{cardinality.v} for details).
Using \L!fsets! and the notation for iterated operators over
finite supports from Sect.~\ref{sec:fsbigop}, the pencil-and-paper
definition of sums over general sets translates directly:

\begin{lstlisting}
Variables (R : realFieldType) (T : choiceType).
Definition fsets S : set (set T) := [set F | finite_set F /\ F `<=` S].

Definition esum (S : set T) (a : T -> \bar R) :=
  ereal_sup [set \sum_(x \in A) a x | A in fsets S].
\end{lstlisting}
The type \L!realFieldType! is one of the numeric types of \mathcomp{} (see Fig.~\ref{fig:numtypes}).
The identifier \L!ereal_sup! corresponds to
the supremum of a set of extended real numbers.
The definition \L!esum! is equipped with the notation \L!\esum_(i in P) f i!
of Table~\ref{tab:iterated}.
It generalizes the notation for the limit of
sequences of extended real numbers of Sect.~\ref{sec:ereal_algebraic}.
As an illustration of the theory of sums over general sets, let us
consider the following partition property:
\begin{align}
J_k \textrm{ pairwise-disjoint} \to (\forall j, j \in \bigcup_{k \in K} J_k \to 0 \leq a_j) \to \nonumber \\
\sum_{i \in \bigcup_{k \in K} J_k} a_i = \sum_{k \in K} \left(\sum_{j \in J_k} a_j\right). \nonumber
\end{align}
Here follows the corresponding formal statement, where the hypothesis
about the pairwise-disjointness of the sets $J_k$ is slightly
generalized (see Table~\ref{tab:set} for notations):
\begin{lstlisting}
Lemma esum_bigcup J a : trivIset [set k | a @` J k != [set 0]] J ->
    (forall x, (\bigcup_(k in K) J k) x -> 0 <= a x) ->
  \esum_(i in \bigcup_(k in K) J k) a i =
  \esum_(k in K) \esum_(j in J k) a j.
\end{lstlisting}
This property will turn out to be useful when developing the Measure
Extension theorem later in this paper.

\section{Basic Definitions of Measure Theory}
\label{sec:measure_theory}

The main mathematical definitions for measure theory are \salgebra{}
and measure. The goal of the construction of the Lebesgue measure is
to build a function that satisfies the properties of a measure. This
is not trivial because such a function does not exist in general when
the domain is an arbitrary powerset, hence the introduction of
\salgebra{}s.

This section proposes a formalization of the basic definitions of
measure theory using \hb{}~\cite{cohen2020fscd}, a tool that automates
the writing of \newterm{packed classes}~\cite{garillot2009tphols}, a
methodology to build hierarchies of mathematical structures that is used
pervasively in the Mathematical Components project.

\subsection{Overview of \hb{}}
\label{sec:hb_overview}

\hb{} extends \coq{} with commands to define hierarchies of mathematical
structures and functions.
It is designed so that hierarchies can evolve (for example by
splitting a structure into smaller structures) without breaking
existing code.
These commands are compiled to packed classes~\cite{garillot2009tphols},
but the technical details of their implementation in \coq{} (modules,
records, coercions, implicit arguments, canonical structures
instances, notations, etc.) are hidden to the user.

The main concept of \hb{} is the one of \newterm{factory}. This is a record
defined by the command \L!HB.factory! that packs a carrier,
operations, and properties. This record usually corresponds to the
standard definition of a mathematical structure. \newterm{Mixins} are
factories used as the default definition for a mathematical structure;
they are defined by the command \L!HB.mixin!.
\newterm{Structures} defined by the command \L!HB.structure! are
essentially sigma-types with a carrier paired with one or more
factories.
A mixin often extends a structure, so it typically takes as
parameters a carrier and other structures.

Factories are instantiated using the command
\L!HB.instance!. Instances are built with an \L!xyz.Build! function
which is automatically generated for each \L!xyz! factory.

A \newterm{builder} is a function that shows that a factory is
sufficient to build one or several mixins.
To add builders, one uses the command \L!HB.builders! that opens a
\coq{} section which starts with postulating a factory instance and
lets the user declare several instances of mixins as builders.

In addition to commands to build hierarchies, \hb{} also checks their
validity by detecting missing interfaces or \newterm{competing
  inheritance paths}~\cite{affeldt2020ijcar}. More than an inheritance
mechanism, \hb{} therefore provides help in the design of hierarchies
of structures.

\subsection{Mathematical Structures for Measure Theory}
\label{sec:math_struct_measure}

A \salgebra{} is a mathematical structure that comprises a set of sets
that contains the empty set, and is stable by complement and by
countable union. It is best defined as a hierarchy of mathematical
structures because more general structures actually play a key
role in the construction by extension of the Lebesgue measure.

\subsubsection{Inheritance Chain from Semiring of Sets}
\label{sec:inheritance_chain}

The hierarchy of mathematical structures for measure theory starts
with \newterm{semirings of sets}.
They are formalized using \hb{} (see Sect.~\ref{sec:hb_overview}) as follows:
\begin{lstlisting}[numbers=left,xleftmargin=3.0ex]
HB.mixin Record isSemiRingOfSets (d : measure_display) T := { 7\label{loc:isSemiRingOfSets}7
  measurable : set (set T) ; 7\label{loc:measurable}7
  measurable0 : measurable set0 ; 7\label{loc:measurable0}7
  measurableI : setI_closed measurable; 7\label{loc:measurableI}7
  semi_measurableD : semi_setD_closed measurable }. 7\label{loc:measurableD}7

#[short(type=semiRingOfSetsType)] 7\label{loc:semiRingOfSetsType}7
HB.structure Definition SemiRingOfSets d := 7\label{loc:SemiRingOfSets}7
  {T of isSemiRingOfSets d T}.
\end{lstlisting}
The declaration of the mixin starts at line \ref{loc:isSemiRingOfSets}.
The parameter~\L!d! is what we call a \newterm{display parameter}.
It can be ignored on a first reading because it is not used in the definition;
it is used implement user-friendly notations as will be explained
in Sect.~\ref{sec:display} where we will have enough material to
demonstrate its use with a concrete example.
Line \ref{loc:measurable} corresponds to the carrier.
A semiring of sets contains the empty set (line \ref{loc:measurable0}).
It is also stable by finite intersection;
this is captured by line~\ref{loc:measurableI}, where \L+setI_closed G+ is formally defined as
\L+forall A B, G A -> G B -> G (A `&` B)+.
At line~\ref{loc:measurableD}, \L+semi_setD_closed G+
means that the relative complement of two sets in \L+G+ can be partitioned into
a finite number of sets in \L+G+:
\begin{lstlisting}
Definition "*semi_setD_closed*" G := forall A B, G A -> G B -> exists D,
 [/\ finite_set D, D `<=` G, A `\` B = \bigcup_(X in D) X & trivIset D id].
\end{lstlisting}
The definition of semiring of sets is completed at line~\ref{loc:SemiRingOfSets} by
declaring the structure (as explained in Sect.~\ref{sec:hb_overview}) and providing a conventional
notation for the corresponding type (line~\ref{loc:semiRingOfSetsType}).
Hereafter, we call \newterm{measurable sets} the sets that form a semiring of sets.

A \newterm{ring of sets} is a non-empty set of
sets that is closed under union and difference.
It can be defined by extending a semiring of sets
with the axiom that it is stable by finite union.
Its interface can be defined using \hb{} as follows:
\begin{lstlisting}[numbers=left,xleftmargin=3.0ex]
HB.mixin Record SemiRingOfSets_isRingOfSets d T
    of SemiRingOfSets d T := {7\label{loc:ringextension}7
  measurableU : @setU_closed T measurable }. 7\label{loc:measurableU}7 7\label{loc:the}7

#[short(type=ringOfSetsType)]
HB.structure Definition RingOfSets d := 7\label{loc:RingOfSets}7
  {T of SemiRingOfSets_isRingOfSets d T & SemiRingOfSets d T 7\label{loc:extends}7}.
\end{lstlisting}
This declaration provides a new mixin that extends the mixin for
semiring of sets (note the \L!of! declaration at line~\ref{loc:ringextension}).
At line~\ref{loc:measurableU},
the expression \L+setU_closed G+ means that the class \L+G+ is stable by finite
unions and is formally defined as %
\L+forall A B, G A -> G B -> G (A `|` B)+.  The modifier \L!@! at line~\ref{loc:the} is
\coq{} syntax to enforce the explicit input of implicit arguments.
%
%
The corresponding structure is declared at line~\ref{loc:RingOfSets}
where it is marked as satisfying the mixin \L!SemiRingOfSets_isRingOfSets!
and extending the structure of semiring of sets \L!SemiRingOfSets! (line \ref{loc:extends}).

\begin{figure}[htbp]
\centering
\includegraphics[width=2.5cm]{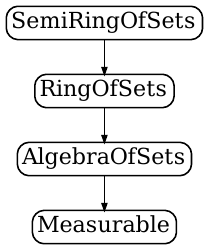}
\caption{Inheritance chain from semiring of sets to \salgebra{}'s}
\label{fig:inheritance_chain}
\end{figure}

An \newterm{algebra of sets} is a set of sets that contains the empty
set and is stable by (finite) union and complement.  Algebras of sets
are defined as extending rings of sets with the axiom that the full
set belongs to the set of measurable sets.  The \hb{} declaration is similar to
the one of semiring of sets and ring of sets:
\begin{lstlisting}
HB.mixin Record RingOfSets_isAlgebraOfSets d T of RingOfSets d T :=
  { measurableT : measurable [set: T] }.

#[short(type=algebraOfSetsType)]
HB.structure Definition AlgebraOfSets :=
  {T of RingOfSets_isAlgebraOfSets T & RingOfSets d T}.
\end{lstlisting}

Finally, \salgebra{}s are defined by extending algebras of sets
with the axiom of stability by countable union:
\begin{lstlisting}
HB.mixin Record Measurable_from_algebraOfSets d T
    of AlgebraOfSets d T :=
  { bigcupT_measurable :
    forall F, (forall k, measurable (F k)) -> measurable (\bigcup_k (F k)) }.

#[short(type=measurableType)]
HB.structure Definition Measurable :=
  {T of Measurable_from_algebraOfSets d T & AlgebraOfSets d T}.
\end{lstlisting}
These definitions form an inheritance chain
(Fig.~\ref{fig:inheritance_chain}), so that \salgebra{}s are also
algebras of sets, which are also rings of sets, and therefore semirings
of sets.

\subsubsection{Direct Definition of Mesurable Spaces}
\label{sec:measurable_factory}

The set of interfaces provided by the hierarchy of mathematical
structures for measure theory is not the only way to instantiate
structures.  We also provide factories (introduced in Sect.~\ref{sec:hb_overview}).
For example, the following factory provides an alternative interface for
\salgebra{}s:
\begin{lstlisting}
HB.factory Record isMeasurable d T := {
  measurable : set (set T) ;
  measurable0 : measurable set0 ;
  measurableC : forall A, measurable A -> measurable (~` A) ;
  measurable_bigcup :
    forall F, (forall k, measurable (F k)) -> measurable (\bigcup_k (F k)) }.
\end{lstlisting}
It is arguably closer to the textbook definition that we gave at the
beginning of Sect.~\ref{sec:math_struct_measure}. In fact, though it
may seen at first sight that mixins provide the definition of
mathematical structures, we have observed in practice that the
standard textbook definition often ought better be sought in factories
provided afterwards.

\subsection{Generated \salgebra{}s}
\label{sec:generated_salgebra}

The notion of \newterm{generated \salgebra{}} will come in handy to
define the Measure Extension theorem and to develop the theory of
measurable functions.
The generated \salgebra{} \L!<<s D, G >>! is the smallest \salgebra{}
that contains the set of sets \L!G!, such that the complement is taken
w.r.t.\ a set \L!D!.
This is defined using a generic \L+smallest+ predicate:
\begin{lstlisting}
Definition smallest C G := \bigcap_(A in [set M | C M /\ G `<=` M]) A. 7\label{loc:smallest}7
...
Context {T}.
Definition "*sigma_algebra*" D G := [/\ G set0, (forall A, G A -> G (D `\` A)) &
  (forall A : (set T)^nat, (forall n, G (A n)) -> G (\bigcup_k A k))].
...
Notation "'<<s' D , G '>>'" := (smallest ("*sigma_algebra*" D) G).
\end{lstlisting}
Below, the notation \L!<<s G >>! is for the measurable sets of the
\salgebra{} generated from the set of sets \L!G! with complement take
w.r.t.\ the full set.

Note that the definition \L+smallest+ is well-defined (i.e., is indeed
the smallest set in the class \L+C+) whenever the smallest fixpoint of
the class \L+C+ indeed exists.  This is why the definition of a
generated \salgebra{} can also be found
elsewhere~\cite[\sect{4.2}]{boldo2021jar} defined as an inductive
predicate instead.
The choice of using the \L+smallest+ predicate rather than an inductive definition
is for the sake of genericity: we have a unique function symbol
and a common theory to deal with all generated classes
(Dynkin, \salgebra{}, etc.), and since \L+smallest+ itself
is monotonous, we can reduce comparison of generated classes to
the extent of the classes themselves.
However, this has the drawback
that the elimination principle and correctness lemmas
are not automatically proven by \coq{} as they would with the
\L!Inductive! command.

Since the set of sets of type \L!<<s G >>! forms a \salgebra{} we can
equip it with the structure of \L!measurableType! from
Sect.~\ref{sec:inheritance_chain}.
The temptation would be to give a definition to the measurable type
generated by \L!G!. However, for the sake of modularity it is a
common practice in the \mathcomp{} libraries to define a type alias of
\L!T! (of type \L!pointedType!, i.e., it contains at least one element)
with~\L!G! as a phantom argument and provide a inferable instance of
measurable type on this alias.
Hence we introduce a dedicated identifier \L!salgebraType!
(line~\ref{salgebraType} below) to alias \L!T! and a dedicated display
parameter \L!sigma_display! (line~\ref{line:sigma_display}) (remember
from Sect.~\ref{sec:inheritance_chain} that mathematical structures
for measure theory are parameterized by a display parameter to be
explained in Sect.~\ref{sec:display}).
Let us furthermore assume that we are given the proofs \L!sigma_algebra{0,C,_bigcup}! corresponding
to the \salgebra{} properties of a generated \salgebra{}.
To associate to the identifier \L!salgebraType! a structure of \salgebra{}, we
use the \hb{} command \L!HB.instance!.  At line~\ref{line:hbinstancepointed},
it equips \L!salgebraType! with a structure of pointed type (\L+Pointed.on+
replicates the pointed type structure of the underlying type, here \L+T+).
At line~\ref{line:hbinstance}, it is used with the
constructor of the factory \L!isMeasurable! of
Sect.~\ref{sec:measurable_factory}.
The corresponding display appears at line~\ref{line:isMeasurabledisplay} and
the proofs of the \salgebra{} properties appear at lines \ref{line:proofs1}--\ref{line:proofs2}.
\begin{lstlisting}[numbers=left,xleftmargin=3.0ex]
Definition salgebraType {T} (G : set (set T)) := T. 7\label{salgebraType}7
Definition sigma_display {T} : set (set T) -> measure_display. 7\label{line:sigma_display}7
Proof. exact. Qed.

Section g_salgebra_instance.
Variables (T : pointedType) (G : set (set T)).

HB.instance Definition _ := Pointed.on (salgebraType G). 7\label{line:hbinstancepointed}7
HB.instance Definition _ := @isMeasurable.Build 7\label{line:hbinstance}7
  (sigma_display G) 7\label{line:isMeasurabledisplay}7
  (salgebraType G) <<s G >>
  (@sigma_algebra0 _ setT G) (@sigma_algebraC) 7\label{line:proofs1}7
  (@sigma_algebra_bigcup _ setT G). 7\label{line:proofs2}7

End g_salgebra_instance.
\end{lstlisting}

\subsection{Displays for Measurable Types}
\label{sec:display}

We saw in the previous sections that the structures for measure theory are
parameterized by a display parameter. Its purpose is to disambiguate
the printing of expressions of the (input) form \L!measurable A!. This is useful when several
of them appear in the same local context or when \L!A! does not provide
enough information to infer the right measurable type.

More concretely, let us consider the basic case of a measurable type~\L!T! with display~\L!d!
(e.g., \L!T : ringOfSetsType d!). To assert that a set \L!A : set T! is measurable,
one can always write \L!measurable A!.
Yet, the display mechanism is such that \coq{} prints
back \L!d.-measurable A!. This is achieved by providing a type
for displays (line~\ref{line:measure_display}) and a notation
(line~\ref{line:dmeasurable}):
\begin{lstlisting}[numbers=left,xleftmargin=3.0ex]
Inductive measure_display := default_measure_display. 7\label{line:measure_display}7

Declare Scope measure_display_scope.
Delimit Scope measure_display_scope with mdisp.

Notation "d .-measurable" := (@measurable d%mdisp). 7\label{line:dmeasurable}7
\end{lstlisting}

The display mechanism can be used to disambiguate expressions. Let us consider
the case of generated \salgebra{}'s. We saw that the display for generated \salgebra{}'s
is parameterized by the generator set (\L!sigma_display! in the previous section---Sect.~\ref{sec:generated_salgebra}).
We can therefore introduce a notation
\L!G.-sigma! for the display associated with the generator set \L!G! and
a notation \L!G.-sigma.-measurable! for the measurable sets of the \salgebra{}
generated by \L!G!:
\begin{lstlisting}
Notation "G .-sigma" := (sigma_display G) : measure_display_scope.
Notation "G .-sigma.-measurable" :=
  (measurable : set (set (salgebraType G))) : classical_set_scope.
\end{lstlisting}
For example, we can use these notations to regard the empty set
\L!set0!
as a member of the \salgebra{} generated by any set \L!G!:
\begin{lstlisting}
Goal forall (T : pointedType) (G : set (set T)), G.-sigma.-measurable set0.
Proof. by move=> T G; exact: measurable0. Qed.
\end{lstlisting}
In comparison, the input \L!measurable set0! would not type check, because
\L+T+ does not have a default instance to give a meaning to \L+measurable+.

\subsection{Functions on Classes of Sets}
\label{sec:functions_classes_sets}

There are several notions of functions from classes of
sets to the real numbers (or, implicitly, extended reals)
which fall under the umbrella name
of ``measure''. In the literature, they are
named \newterm{contents} (a.k.a.\ additive measures), premeasures,
outer measures, \ssubadditive\ measures, and
\sadditive\ measures (a.k.a.\ measures).
We define predicates for all of these notions, but we only define
structures for the three most useful: contents, measures, and outer measures.

\subsubsection{Contents}
\label{sec:content}

A \newterm{content} (or an additive measure)~$\mu$ is a non-negative function defined
over a semiring of sets such that the measure of the empty set is $0$
and such that $\mu(\cup_{k=1}^n F_k) = \sum_{k=1}^n \mu(F_k)$ for a finite number of
pairwise-disjoint measurable sets~$F$.
We first provide a definition for the latter condition:
\begin{lstlisting}
Definition semi_additive mu := forall F n, (forall k, measurable (F k)) ->
  trivIset setT F -> measurable (\big[setU/set0]_(k < n) F k) ->
  mu (\big[setU/set0]_(k < n) F k) = \sum_(k < n) mu (F k).
\end{lstlisting}
The pairwise-disjointness of sets is captured by the generic predicate
\L!trivIset! (Table~\ref{tab:set}).
Asking $\cup_{k=1}^n F_k$ to be measurable is superfluous when taken on a
ring of sets.
Contents are eventually defined by the following mixin and
structure:
\begin{lstlisting}
HB.mixin Record isContent d (T : semiRingOfSetsType d)
    (R : numFieldType) (mu : set T -> \bar R) := {
  measure_ge0 : forall x, 0 <= mu x ;
  measure_semi_additive : semi_additive mu }.
HB.structure Definition Content d (T : semiRingOfSetsType d)
    (R : numFieldType) := { mu & isContent d R T mu }.
\end{lstlisting}
See Fig.~\ref{fig:numtypes} for the type \L!numFieldType!.
In the \coq{} code, the type of contents is denoted by
\L!{content set T -> \bar R}!.

An essential property of contents is that they can be extended
from a semiring of sets~$\mathcal{S}$ to its generated ring of
sets~$R(\mathcal{S})$. We can define the latter similarly to how
we defined generated \salgebra{}s in Sect.~\ref{sec:generated_salgebra}:
\begin{lstlisting}
Definition setring G := [/\ G set0, setU_closed G & setDI_closed G].
Notation "'<<r' G '>>'" := (smallest setring G).
\end{lstlisting}
The predicate \L!setDI_closed! is for stability by set difference
and is defined by \L!forall A B, G A -> G B -> G (A `\` B)!.

A generated ring of sets can be equipped with a canonical structure of
ring of sets. It happens that the measurable sets of these generated rings of sets
can in fact be expressed as the finite disjoint unions of (non-empty) sets
from the original semiring of sets~$\mathcal{S}$ (\L!rT! in the lemma below indicates
sets from the generated ring of sets):
\begin{lstlisting}
Lemma "*ring_finite_set*" (A : set rT) : measurable A -> exists B : set (set T),
  [/\ finite_set B,
      (forall X, B X -> X !=set0),
      trivIset B id,
      (forall X, X \in B -> measurable X) &
      A = \bigcup_(X in B) X].
\end{lstlisting}
Thanks to this lemma, we can make this decomposition explicit by the
following function~\L+decomp+, which given a set $A$ in~$R(\mathcal{S})$
returns a finite set of sets in~$\mathcal{S}$ that cover $A$:
\begin{lstlisting}
Definition decomp (A : set rT) : set (set T) :=
  if A == set0 then [set set0]
  else if pselect (measurable A) is left mA then
    projT1 (cid ("*ring_finite_set*" mA))
  else [set A].
\end{lstlisting}
The function \L!decomp! is written in an idiomatic way to retrieve in
\coq{} a witness from an existential proof. The identifier \L!pselect!
comes from \analysis{} and is a strong version of the law of excluded
middle~\cite[Sect.~5.2]{cohen2018jfr}; \L!cid! is the axiom of
constructive indefinite description.

Using \L!decomp!, we can extend the function over the original semiring
of sets by summing the components:
\begin{lstlisting}
Definition measure (R : numDomainType) (mu : set T -> \bar R)
  (A : set rT) : \bar R := \sum_(X \in decomp A) mu X.
\end{lstlisting}
We thus have a \L+measure mu+ function for all functions \L+mu+, which is equal
to \L+mu+ on the sets of the semiring of sets where \L+mu+ is defined, and
which is a content on the generated ring of sets when \L+mu+ is a content
(section \L!content! in~\accompanying{measure.v}), and which is
\sadditive{} if \L+mu+ is \ssubadditive{} (lemma
\L+ring_semi_sigma_additive+). Furthermore, using
the latter fact we prove that when \L+mu+ is \ssubadditive on a semiring of sets,
it is in fact \sadditive{} (lemma \L!semiring_sigma_additive!).

\subsubsection{Measures}
\label{sec:measure}

A \newterm{measure}~$\mu$ is defined similarly to a content.
The difference is the additivity axiom: it is such that
$\mu(\cup_k F_k) = \sum_k \mu(F_k)$ for any sequence~$F$
of pairwise-disjoint measurable sets.
We provide a definition for the latter condition, but
generalizing it for semirings of sets by requiring the
union $\cup_k F_k$ to be measurable as a precondition, thus merging
the notions of measure and premeasure into one:
\begin{lstlisting}
Definition semi_sigma_additive mu := forall F, (forall k, measurable (F k)) ->
  trivIset setT F -> measurable (\bigcup_k F k) ->
  (fun n => \sum_(k < n) mu (F k)) --> mu (\bigcup_k F k).
\end{lstlisting}
The notation \L!f --> l! is a notation for convergence of functions
that comes from \analysis{}~\cite{cohen2018jfr}. In particular, when
\L!f --> l! holds, we have \L!lim f = l! using the \L!lim! notation of
Sect.~\ref{sec:ereal} (provided that the range of the function \L!f! is
a separated space, which is the case for the functions considered in this section).
Note that in the definition above the precondition \L!measurable (\bigcup_k F k)!
holds unconditionally whenever we know that the underlying type is a \salgebra{}.

We use this definition to define the mixin corresponding to measures,
which extends the one for contents:
\begin{lstlisting}
HB.mixin Record Content_isMeasure d (T : semiRingOfSetsType d)
    (R : numFieldType) mu of isContent d R T mu := {
  measure_semi_sigma_additive : semi_sigma_additive mu }.

#[short(type=measure)]
HB.structure Definition Measure d (T : semiRingOfSetsType d)
    (R : numFieldType) :=
  { mu of Content_isMeasure d T mu & Content d mu }.
\end{lstlisting}

In practice, to construct a measure, one would rather use the following factory
(we introduced the notion of factory in Sect.~\ref{sec:measurable_factory})
whose interface is closer to the textbook definition of measure:
\begin{lstlisting}
HB.factory Record isMeasure d (T : semiRingOfSetsType d)
   (R : realFieldType) (mu : set T -> \bar R) := {
  measure0 : mu set0 = 0 ;
  measure_ge0 : forall x, 0 <= mu x ;
  measure_semi_sigma_additive : semi_sigma_additive mu }.
\end{lstlisting}
The notation \L!{measure set T -> \bar R}! corresponds to the type of measures.

\subsubsection{Outer Measures}
\label{sec:outer_measure}

\newterm{Outer measures} are the object of study of the measure extension theorems.
Contrarily to a measure, an outer measure \L+mu+ is \ssubadditive{} on the full powerset
rather than on a specific class of sets.
\begin{lstlisting}
Definition sigma_subadditive
    (R : numFieldType) (T : Type) (mu : set T -> \bar R) :=
  forall (F : (set T)^nat), mu (\bigcup_n (F n)) <= \sum_(n <oo) mu (F n).
\end{lstlisting}
Compared to \sadditivity{}, in \ssubadditivity{} the relation
between the measure of the countable union and the sum of the measures
is an inequality, there are no conditions on the sequence of sets, and
the support type need not be a \salgebra{}.
Like for contents and measures
(Sections~\ref{sec:content} and~\ref{sec:measure}), we encode
an outer measure as a \hb{} mixin:
\begin{lstlisting}
HB.mixin Record isOuterMeasure
    (R : numFieldType) (T : Type) (mu : set T -> \bar R) := {
  outer_measure0 : mu set0 = 0 ;
  outer_measure_ge0 : forall x, 0 <= mu x ;
  le_outer_measure : {homo mu : A B / A `<=` B >-> A <= B} ;
  outer_measure_sigma_subadditive : sigma_subadditive mu }.
\end{lstlisting}
The notation \L!{homo f : x y / r x y >-> s x y}!  is a generic \mathcomp{}
notation for homomorphisms \L!f! with respect to the relations \L!r!
and \L!s!.
The type of outer measures comes with the notation \L!{outer_measure set T -> \bar R}!.


\section{Measure Extension}
\label{sec:extension}

A standard approach to the construction of measures is to extend a
function over a semiring of sets, a ring of sets, or an algebra of
sets to a measure over an enclosing \salgebra{}.
These extension theorems are known under different names
(Carath\'eodory/Carath\'eodory-Fréchet/Carath\'eodory-Hopf/Hopf/Hahn/Hahn-Kolmogorov/etc.\
extension theorems).
In the following, we explain the formalization of a version starting
from semiring of sets and refer to it as the Measure Extension
theorem.

As in the textbooks we follow~\cite{klenke2014,liintegration}, we
decompose the Measure Extension theorem in reusable constructions and
lemmas. The first, which we refer to as the \newterm{outer measure
  construction}, extends a non-negative function $\mu$ such that
$\mu(\emptyset) = 0$ over a semiring of sets $\mathcal{S}$ to an outer
measure (Sect.~\ref{sec:caratheodory1}).
This is then shown to be a measure over the \salgebra{} of
\newterm{Carath\'eodory-measurable sets} (Sect.~\ref{sec:caratheodory2}). When restricted
to this \salgebra{}, we call it the \newterm{Carath\'eodory measure}.
Now, if $\mu$ was a \ssubadditive{} content on $\mathcal{S}$, the
\salgebra{} of Carath\'eodory-measurable sets contains the \salgebra{}
generated by~$\mathcal{S}$, and the Carath\'eodory measure is uniquely
determined on it, by the values of $\mu$ on $\mathcal{S}$
(Sect.~\ref{sec:hahn_extension}).

\subsection{Outer Measure Construction}
\label{sec:caratheodory1}

The first part of the Measure Extension theorem builds an outer
measure (Sect.~\ref{sec:outer_measure}) given a function defined over
a semiring of sets.
In textbooks it is often stated in a weaker form starting from a ring
of sets or an algebra of sets.
The outer measure in question
is more precisely defined as the infimum of the measures of covers, i.e.,
$\inf_F\left\{ \sum_{k=0}^\infty \mu(F_k) \,\text{\textbar}\, (\forall k, \measurable{F_k} ) \land X \subseteq \bigcup_k F_k\right\}.$
The definition of these coverable measures translates directly in \analysis{}:
\begin{lstlisting}
Definition measurable_cover X :=
  [set F | (forall k, measurable (F k)) /\ X `<=` \bigcup_k (F k)].
\end{lstlisting}
We use \L!measurable_cover! to define the desired outer measure:
\begin{lstlisting}
Context d (T : semiRingOfSetsType d) (R : realType).
Variable mu : set T -> \bar R.
Definition "*mu_ext*" (X : set T) : \bar R :=
  ereal_inf [set \sum_(k <oo) mu (F k) | F in measurable_cover X].
\end{lstlisting}
The identifier \L!ereal_inf! corresponds to
the infimum of a set of extended real numbers.
In the following, \L!mu_ext mu! is noted \L!mu^*!.

The difficulty to show that \L!mu^*! is an outer measure is to show
that it is \ssubadditive{} (remember that we are working under the
hypotheses that \L!mu set0 = 0! and that \L!mu! is non-negative).  A
typical textbook proof~\cite[\sect{X.1}]{liintegration}\cite[Lemma
1.47]{klenke2014} translates to a proof script of 54 lines of code
(lemma \L!mu_ext_sigma_subadditive!, \accompanying{measure.v}).
The main technical point is the use of sums over general sets. Precisely, in the course of
proving \ssubadditivity{}, we run into a subgoal of the following shape ($\mu^*$
is the outer measure under construction):
$$
\mu^*(\cup_i F_i) \leq \sum_i^\infty \left(\mu^*(F_i) + \frac{\varepsilon}{2^i}\right).
$$
The proof goes on by showing
$
\mu^*(\cup_i F_i) \leq \sum_{i,j} \mu(G_i j) \leq \sum_i \sum_j \mu(G_i \, j)
$
for some well-chosen~$G$,
such that
$
F_i \subseteq \cup_j G_i j
$
and
$
\sum_j \mu(F_i j) \leq \mu^*(F_i) + \varepsilon/2^i
$.
This proof can be completed with the partition property using sums over general sets
from Sect.~\ref{sec:esum}.

Coming back to \L!mu^*!, we also show that it coincides with the
input measure~\L!mu! (lemma \L!measurable_mu_extE! in \accompanying{measure.v}).

\subsection{From an Outer Measure to a Measure}
\label{sec:caratheodory2}

The second part of the Measure Extension theorem builds, given an outer
measure, a \salgebra{} and a measure over it.  The resulting
\salgebra{} is formed of Carath\'eodory measurable sets,
i.e., sets~$A$ such that
$\forall X, \mu^*(X)=\mu^*(X\cap A) + \mu^*(X\cap\bar{A})$ where
$\mu^*$ is an outer measure.
Hereafter, the set of Carath\'eodory measurable sets for an outer measure \L!mu!
will appear as the notation \L!mu.-caratheodory!.

Given our newly developed theory of sequences of extended real numbers
(Sect.~\ref{sec:ereal_seq}), proving, for an outer measure \L!mu!,
that \L!mu.-caratheodory! is actually a \salgebra{} is essentially a
translation of standard pencil-and-paper proofs (see lemmas
\L!caratheodory_measurable_{set0,setC,bigcup}! in \accompanying{measure.v}).
Hereafter, the \salgebra{} of Carath\'eodory measurable sets is denoted by \L!mu.-cara.-measurable!
(this notation is implemented using the display mechanism
explained in Sect.~\ref{sec:display}).

Similarly, proving that the restriction of the outer measure \L!mu! to
the \salgebra{} \L!mu.-cara.-measurable!  is a measure is also
essentially a direct translation of standard pencil-and-paper proofs (see
lemmas \L!caratheodory_measure{0,_ge0,_sigma_additive}!).

Finally, we formally prove a number of properties about the
resulting measure, in particular that it is \newterm{complete}, i.e.,
\newterm{negligible sets} are measurable. Let~\L!T! be a semiring of sets
and \L!R! be a \L!realFieldType!.
A set $N$ is negligible for $\mu$ when there exists a measurable set
$A$ such that $\mu(A)=0$ and $N \subseteq A$:
\begin{lstlisting}
Definition negligible (mu : set T -> \bar R) N :=
  exists A, [/\ measurable A, mu A = 0 & N `<=` A].
\end{lstlisting}
Let \L!mu.-negligible X! be a notation for \L!X! is negligible.
The formal definition of a complete measure follows:
\begin{lstlisting}
Definition measure_is_complete (mu : set T -> \bar R) :=
  mu.-negligible `<=` measurable.
\end{lstlisting}

\subsection{The Measure Extension Theorem}
\label{sec:hahn_extension}

Finally, we show that a measure over a semiring of sets can be extended to
a measure over a \salgebra{} that contains all the measurable sets of
the smallest \salgebra{} containing the semiring of sets.
We place ourselves in the following context:
\begin{lstlisting}
Context d (T : semiRingOfSetsType d) (R : realType).
Variable mu : {measure set T -> \bar R}.
\end{lstlisting}
In this context, we can build an outer measure \L!mu^*! using the
results of Sect.~\ref{sec:caratheodory1} and its \salgebra{}
\L!mu^*.-cara.-measurable! using the results of
Sect.~\ref{sec:caratheodory2}.
We can show that this \salgebra{} contains all the measurable sets
generated from the semiring of sets:
\begin{lstlisting}
Lemma sub_caratheodory :
  (d.-measurable).-sigma.-measurable `<=` mu^*.-cara.-measurable.
\end{lstlisting}
Recall from Sect.~\ref{sec:display} that \L!G.-sigma.-measurable!
corresponds to the \salgebra{} generated from \L!G!  and that in our
context \L!d.-measurable! corresponds to the measurable sets of the
semiring of sets~\L!T!. As for \L!m.-cara.-measurable!, we saw in
Sect.~\ref{sec:caratheodory2} that it corresponds to the \salgebra{}
of Carath\'eodory measurable sets for the outer measure~\L!m!.

We use this last fact to build a measure over the \salgebra{}
generated from the semiring of sets: this is the final result
of the Measure Extension Theorem
(recall from Sect.~\ref{sec:generated_salgebra} that \L!salgebraType G!
is the measurable type generated by~\L!G!):
\begin{lstlisting}
Let I := salgebraType (@measurable _ T).
Let measure_extension : set I -> \bar R := mu^*.

HB.instance Definition _ := isMeasure.Build _ _ _ measure_extension
  measure_extension0 measure_extension_ge0
  measure_extension_semi_sigma_additive.
\end{lstlisting}
The proofs \L!measure_extension{0,ge0,_sigma_additive}! correspond to the
properties of a measure as explained in Sect.~\ref{sec:measure}.
See \accompanying{measure.v} for details.

Furthermore, we prove that the measure extension is unique.
This requires to prove beforehand the uniqueness
of measures \cite[lemma \L+measure_unique+]{analysis}.
We use monotone classes for that
purpose~\cite[\sect{V.2.1}]{liintegration}.
This can also be proved using the equivalent notion of Dynkin systems
(as mentioned in~\cite{holzl2011itp}, which we also formalized in
\accompanying{measure.v}).
The uniqueness of measure extension is under the condition that the
measure is \sfinite{}, i.e., the full set can be covered by a
countable union of sets of finite measure:
\begin{lstlisting}
Definition sigma_finite (A : set T) (mu : set T -> \bar R) :=
  exists2 F : (set T)^nat, A = \bigcup_(i : nat) F i &
    forall i, measurable (F i) /\ mu (F i) < +oo.
\end{lstlisting}
When this holds for the measure of the measure extension, any
other measure \L!mu'! that coincides with \L!mu!  on the original
semiring of sets also coincides with the measure extension over the
generated \salgebra{}:
\begin{lstlisting}
Lemma measure_extension_unique : sigma_finite [set: T] mu ->
  (forall mu' : {measure set I -> \bar R},
    (forall X, d.-measurable X -> mu X = mu' X) ->
    (forall X, (d.-measurable).-sigma.-measurable X ->
      measure_extension X = mu' X)).
\end{lstlisting}

Since \sfinite{} measures are actually pervasive in measure theory, we
introduce, we extend using \hb{} the hierarchy of structures for
contents and measures of Sections~\ref{sec:content}
and~\ref{sec:measure} with a structure \L!SigmaFiniteContent! for
contents that are \sfinite{} and a structure \L!SigmaFiniteMeasure!
for measures that are \sfinite.
In particular, hereafter, \L!{sigma_finite_measure set T -> \bar R}!
is a notation for the type of \sfinite{} measures.

\section{Construction of the Lebesgue Measure over a Semiring of Sets}
\label{sec:lebesgue_measure}

In this section, we explain how we derive the Lebesgue measure from
the semiring of sets of intervals of the form $]a, b]$ using the
measure extension from the previous section.

\subsection{The semiring of sets of Intervals}

In \mathcomp{}, the type \L!interval R!, where \L!R! is typically an
ordered type, is defined as the pairs of bounds of type \L!itv_bound!:
\begin{lstlisting}
Variant itv_bound (T : Type) : Type :=
  BSide : bool -> T -> itv_bound T | BInfty : bool -> itv_bound T.
Variant interval (T : Type) := Interval of itv_bound T & itv_bound T.
\end{lstlisting}
The constructor \L!BSide! is for open or closed bounds,
\L!BInfty! is for infinite bounds. How the boolean parameter distinguishes
between open and closed bounds is better explained with illustrations.
For example, the left bounds of the intervals \L!`[x, +oo[! and
\L!`]x, +oo[! are respectively \L!BSide true x! and \L!BSide false x!,
while the right bound of the interval \L!`]-oo, x[! is \L!BSide true x!.
This type allows for the statements of generic lemmas about intervals,
when they happen to hold independently of whether a bound is
open or closed.

Let us define a type alias for \L!R! of type \L!realType! and the
following set of open-closed intervals:
\begin{lstlisting}
Definition ocitv_type : Type := R.
Definition ocitv := [set `]x.1, x.2]%classic | x in [set: R * R]].
\end{lstlisting}
This set forms a semiring of sets. Indeed, it contains
\L!set0!, it is closed under finite intersection, and it satisfies the
\L!semi_setD_closed! predicate from Sect.~\ref{sec:inheritance_chain}
(proofs \L!ocitv{0,I,D}! below):
\begin{lstlisting}
Definition ocitv_display : Type -> measure_display. Proof. exact. Qed.
HB.instance Definition _ := Pointed.on ocitv_type.
HB.instance Definition _ :=
  @isSemiRingOfSets.Build (ocitv_display R) ocitv_type
    ocitv ocitv0 ocitvI ocitvD.
\end{lstlisting}

\subsection{Construction of the Lebesgue Measure}
\label{sec:semiring_lebesgue_instance}

The length of an interval is defined by subtracting its left bound
from its right bound. For the sake of generality, this is formally
defined over arbitrary sets for which we take the hull using \L!Rhull!
(see \accompanying{normedtype.v} for the definition of \L!Rhull!):
\begin{lstlisting}
Definition hlength {R : realType} (A : set R) : \bar R :=
  let i := Rhull A in i.2 - i.1.
\end{lstlisting}

Now, the function \L!hlength! is a content on the semiring of sets of intervals.
Indeed, it is non-negative
(proof \L!hlength_ge0! below), and, more importantly,
it is additive over \L!ocitv!:
\begin{lstlisting}
Lemma hlength_semi_additive : semi_additive hlength.
Proof. (* see 7{\color{myred}{\cite{analysis}}}7 *) Qed.
HB.instance Definition _ :=
  isContent.Build R _ hlength hlength_ge0 hlength_semi_additive.
\end{lstlisting}

Moreover, \L!hlength! is also \ssubadditive{}
over \L!ocitv!, and hence a measure:
\begin{lstlisting}
Lemma hlength_sigma_sub_additive : sigma_sub_additive hlength.
Proof. (* see 7{\color{myred}{\cite{analysis}}7 *) Qed.
HB.instance Definition _ := Content_SubSigmaAdditive_isMeasure.Build
  _ _ _ hlength hlength_sigma_sub_additive.
\end{lstlisting}
We obtain the Lebesgue measure as an application of the measure
extension from Sect.~\ref{sec:hahn_extension}. More precisely,
we use the generic definition \L!measure_extension! to define the function
corresponding to the Lebesgue measure, and it is directly a measure from Sect.~\ref{sec:hahn_extension}.
\begin{lstlisting}
Definition lebesgue_measure := measure_extension hlength.
HB.instance Definition _ := Measure.on lebesgue_measure.
\end{lstlisting}

The above construction provides a unique measure that applies to a
\salgebra{} generated from open-closed intervals (remember the use of
\L!salgebraType! in Sect.~\ref{sec:hahn_extension}), which include the
Borel sets: this is the definition of the Lebesgue measure.

We have not introduced an explicit definition for Borel sets but their
\salgebra{}\/ can be denoted by \L!ocitv.-sigma.-measurable! which is a
notation that combines the definition \L!ocitv!  of the set of
open-closed intervals and the notation \L!.-sigma.-measurable!  that
we introduced in Sect.~\ref{sec:display}. This \salgebra{} can easily
be shown to be the same as the one generated by open intervals:
\begin{lstlisting}
Module RGenOpens.
Section rgenopens.
Variable R : realType.
Definition G := [set A | exists x y, A = `]x, y[%classic].
Lemma measurableE :
  (@ocitv R).-sigma.-measurable = G.-sigma.-measurable.
Proof. (* see 7{\color{myred}{\cite{analysis}}7 *) Qed.
End RGenOpens.
\end{lstlisting}
Similarly, it can be shown to be the same as the one generated by open
rays, etc. Furthermore, it can also be easily extended to a
\salgebra{} over extended real numbers.  These facts (whose formal
proofs can be found in~\accompanying{lebesgue_measure.v}) are useful
to establish the properties of measurable functions in the next
section.

\section{Construction of the Lebesgue Integral}
\label{sec:lebesgue_integral}

We now show that the infrastructure we have developed for the Lebesgue
measure can be used to develop the theory of the Lebesgue integral up
to Fubini's theorem, which covers the typical set of properties that
demonstrate the usefulness of such a formalization.
This experiment improves in particular on related work in \coq{} by
providing theorems for functions that are not necessary non-negative
and that are extended-real valued, and also be experimenting with
simpler encodings, in particular the one of simple functions.
Hereafter, we shorten code snippets with the following convention:
\L!T! has type \L!measurableType d! for some display parameter \L!d!,
\L!R! has type \L!realType!, and
\L!mu! is a measure of type \L!{measure set T -> \bar R}!.

\subsection{Mesurable Functions}

Ultimately, the Lebesgue integral is about \newterm{measurable
  functions}.  A function is measurable when any preimage of a
measurable set is measurable.  We defined it for functions with
domain~\L!D! as follows:
\begin{lstlisting}
Definition measurable_fun d d' (T : measurableType d)
    (U : measurableType d') (D : set T) (f : T -> U) :=
  measurable D -> forall Y, measurable Y -> measurable (D `&` f @^-1` Y).
\end{lstlisting}
Note that when in the above definition \L!T! or \L!U! are actually
\L!R! or \L!\bar R! with \L!R : realType!, a concrete instance of
\salgebra{} need to have been declared beforehand as explained in
Sect.~\ref{sec:semiring_lebesgue_instance}.

\def\indic#1{\textbf{1}_{#1}}

\subsection{Simple Functions}
\label{sec:simple_function}

\begin{figure}[htbp]
\centering
\includegraphics[width=3.5cm]{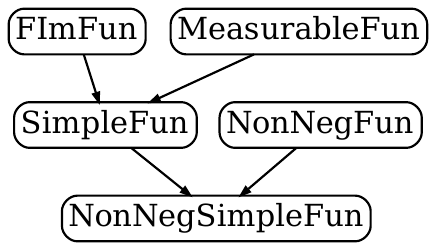}
\caption{Definition of non-negative simple functions}
\label{fig:nnsfun}
\end{figure}

The construction of the Lebesgue integral starts with simple functions.
A \newterm{simple function} $f$ is typically defined by a sequence of
pairwise-disjoint and measurable sets $A_0, \ldots, A_{n-1}$ and a
sequence of elements $a_0, \ldots, a_{n-1}$ such that
$f(x) = \sum_{k=0}^{n-1} a_k\indic{A_k}(x)$.
It might be tempting (in particular for a computer scientist) to
encode this definition using lists to represent the range of simple
functions.
This actually turns out to be detrimental to formalization (see
Sect.~\ref{sec:related_work}).
Instead, we strive for modularity by obtaining simple functions from
even more basic functions. For that purpose, we again put
\hb{} to good use. We first define functions with a finite
image (notation \L!{fimfun T >-> R}!):
\begin{lstlisting}
HB.mixin Record FiniteImage aT rT (f : aT -> rT) :=
  {fimfunP : finite_set (range f)}.
HB.structure Definition FImFun aT rT := {f of @FiniteImage aT rT f}.
\end{lstlisting}
We then package measurable functions (notation \L!{mfun T >-> R}!):
\begin{lstlisting}
HB.mixin Record isMeasurableFun
    d (aT : measurableType d) (rT : realType) (f : aT -> rT) :=
  {measurable_funP : measurable_fun setT f}.
HB.structure Definition MeasurableFun aT rT :=
  {f of @isMeasurableFun aT rT f}.
\end{lstlisting}
As a consequence, simple functions (notation \L!{sfun T >-> R}!) can
be defined by combining both functions with a finite image and measurable functions:
\begin{lstlisting}
HB.structure Definition SimpleFun
    d (aT : measurableType d) (rT : realType) :=
  {f of @isMeasurableFun d aT rT f & @FiniteImage aT rT f}.
\end{lstlisting}
Similarly, we introduce non-negative functions (notation \L!{nnfun T >-> R}!)
and define non-negative simple functions (notation \L!{nnsfun T >-> R}!)
resulting in the hierarchy displayed in Fig.~\ref{fig:nnsfun}.

The introduction for the above collection of types is a fertile ground
for the formalization of the properties of simple functions. We show
in particular that simple functions form a ring structure
(a \L!comRingType! in \mathcomp{}'s parlance) and thus that they can be
combined accordingly (see \L!Section comring! in \accompanying{lebesgue_integral.v}).

Among all the simple functions, indicator functions \L!indic A! (notation \L!\1_A!,
where \L!A! is a set) are of particular interest because they are used
pervasively in the theory of integration:
\begin{lstlisting}
Definition indic {T} {R : ringType} (A : set T) (x : T) : R :=
  (x \in A)%:R.
\end{lstlisting}
(\L+
In particular, any function with a finite image (and thus any simple
function) is a linear combination of indicator functions:
\begin{lstlisting}
Lemma fimfunE T (R : ringType) (f : {fimfun T >-> R}) x :
  f x = \sum_(y \in range f) (y * \1_(f @^-1` [set y]) x).
\end{lstlisting}
This fact is instrumental in proofs using the monotone convergence theorem,
such as Fubini's theorem (Sect.~\ref{sec:dct_fubini}).

\subsection{The Integral of Simple Functions}
\label{sec:sintegral}

The integral of a simple function is the sum of its images multiplied
by the measure of the associated preimage.
In textbooks, the corresponding formula can be written in two ways.
One can make explicit the finite image of the simple function and sum
w.r.t.\ the indices, i.e., as $\sum_{k=0}^{n-1} a_k\mu(A_k)$ using the
notations from the previous section and some measure $\mu$.
Since the image of a simple function is finite,
one can alternatively use sums over finite supports
(Sect.~\ref{sec:fsbigop}) and write:
$\sum_{x \in \mathbb{R}} x \, \mu(f^{-1}\{x\})$.
From the viewpoint of formalization, the former reveals implementation
details while the latter is more compact and allows for the following
simple definition of the integral of simple functions:
\begin{lstlisting}
Variables (T : Type) (R : numDomainType) (mu : set T -> \bar R).
Variable (f : T -> R).
Definition sintegral :=
  \sum_(x \in [set: R]) x%:E * mu (f @^-1` [set x]).
\end{lstlisting}
See Fig.~\ref{fig:numtypes} for \L!numDomainType!.
The development of the properties of the integral of simple functions
goes on by establishing the properties of the integral of non-negative
simple functions such as semi-linearity, monotonicity, etc.  Among
them, the fact that the integral of the sum of simple functions is the
sum of the integrals is the most technical result. Yet, it can be
proved within 23 lines of script using generic properties of sums over
finite supports (see \L!sintegralD!\footnote{\url{https://github.com/math-comp/analysis/blob/7d4ed9cf0e32f6be5b50c092cc8d93a21ec4dee3/theories/lebesgue_integral.v\#L668}}~\accompanying{lebesgue_integral.v}).

\subsection{Integral of Measurable Functions}
\label{sec:integral_measurable_function}

The integral of a measurable function is defined as the difference
between its non-negative part and its non-positive part, both
considered as non-negative functions.
We therefore first temporarily define the integral of a non-negative measurable
function, as the supremum of the integrals of smaller non-negative
simple functions:
\begin{lstlisting}
Let nnintegral f := ereal_sup [set sintegral mu h |
  h in [set h : {nnsfun T >-> R} | forall x, (h x)%:E <= f x]].
\end{lstlisting}
Regarding the definition of the integral of a measurable function, we
make the design choice to have it parameterized with the domain of
integration. For that purpose, we introduce the notation \L!f \_ D!
for the function that behaves as \L!f! over the set \L!D! and $0$
elsewhere. The definition of the integral follows
(notation \L!\int[mu]_(x in D) f x!):
\begin{lstlisting}
Definition integral mu D f (g := f \_ D) :=
  nnintegral mu (g ^\+) - nnintegral mu (g ^\-).
\end{lstlisting}
In the code just above, the notation \L!f ^\+! is for $\lambda x. \max(\texttt{f}(x),0)$ and
the notation \L!f ^\-! is for $\lambda x. \max(-\texttt{f}(x),0)$.

See \accompanying{lebesgue_measure.v} for the development of the
theory of integration as presented in~\cite{liintegration}, and the
next section for two illustrative examples.

\subsection{Dominated Convergence and Fubini's Theorem}
\label{sec:dct_fubini}

The dominated convergence theorem establishes the convergence of a
sequence of integrals of functions $f_n$ given an hypothesis of
pointwise convergence of the functions $f_n$ and an hypothesis of
domination by an integrable function; these two hypotheses are
true ``almost everywhere''.
The standard presentation (e.g.,~\cite[\sect{IV.2}]{liintegration}) is
to first prove the theorem when the hypotheses are
unconditionally true, in which case the proof is essentially a
consequence of Fatou's lemma and of the
linearity properties of the integral.
As for the generalization to hypotheses that are true ``almost everywhere'',
it is almost always only sketched in textbooks.
The complete statement of the dominated convergence
theorem follows. The notation \L!{ae mu, forall x, P x}! means that \L!P!
holds almost everywhere for the measure \L!mu!,
i.e., that the complement of the set defined by \L!P! is negligible as
defined in Sect.~\ref{sec:caratheodory2}.
\begin{lstlisting}
Variables (D : set T) (mD : measurable D).
Variables (f_ : (T -> \bar R)^nat) (f g : T -> \bar R).
Hypothesis mf_ : forall n, measurable_fun D (f_ n).
Hypothesis mf : measurable_fun D f.
Hypothesis f_f : {ae mu, forall x, D x -> f_ ^~ x --> f x}.
Hypothesis ig : mu.-integrable D g.
Hypothesis f_g : {ae mu, forall x n, D x -> `|f_ n x| <= g x}.
Let g_ n x := `|f_ n x - f x|.

Theorem dominated_convergence : [/\ mu.-integrable D f,
  [sequence \int[mu]_(x in D) (g_ n x)]_n --> 0 &
  [sequence \int[mu]_(x in D) (f_ n x)]_n --> \int[mu]_(x in D) (f x) ].
\end{lstlisting}
Note that in this version of the dominated convergence theorem we assume
that \L!f! is measurable; this hypothesis is not needed when \L!mu!
is complete.

Fubini's theorem is a commutation result about integration. It is a
good testbed for a combined formalization of measure and integration
theory because, on the one hand, it requires the construction of the
\newterm{product measure}, and, on the other hand, its proof relies on
several lemmas about integration.
Given two measures \L!m1! and \L!m2! respectively over two measurable
types \L!T1! and \L!T2!, \L!m2! being \sfinite{}, the product measure
\L!m1 \x m2! is defined as \L!\int[m1]_x (m2 \o xsection A) x!  where
\L!xsection A x! is the set of pairs \L!(x, y)! in \L!A!.  In virtue
of the uniqueness of measures (Sect.~\ref{sec:hahn_extension}),
inverting the role of \L!m1! and \L!m2! actually gives rise to the
{\em same} measure.
For the proof of Fubini's theorem, we follow the presentation
by Li~\cite[\sect{V.3}]{liintegration}, which is standard.
The first step is to prove Fubini-Tonelli's theorem, which is
essentially Fubini's theorem for non-negative functions.  The
decomposition of functions with a finite image into a linear
combination of indicator functions (Sect.~\ref{sec:simple_function})
comes in handy to prove Fubini-Tonelli's theorem because the latter is
first established for indicator functions, then for simple
functions, and finally for measurable functions. The second main
ingredient is the monotone convergence theorem~\accompanying{lebesgue_integral.v}.
Fubini's theorem is then essentially an application of
Fubini-Tonelli's theorem:
\begin{lstlisting}
Context d2 d2 (T1 : measurableType d1) (T2 : measurableType d2)
  (R : realType).
Variables (m1 : {sigma_finite_measure set T1 -> \bar R})
          (m2 : {sigma_finite_measure set T2 -> \bar R}).
Variable f : T1 * T2 -> \bar R.
Hypothesis imf : (m1 \x m2).-integrable setT f.

Theorem Fubini :
  \int[m1]_x (\int[m2]_y f (x, y)) = \int[m2]_y (\int[m1]_x f (x, y)).
\end{lstlisting}

\section{Related Work}
\label{sec:related_work}

\paragraph*{About Measure and Integration Theory in Proof Assistants based on Dependent Type Theory}
We are not aware of any formalization of the measure extension theorem
{\em for general semirings of sets}
in a proof assistant based on dependent type theory (neither \coq{} nor \lean).

There is a formalization in \coq{}, based on the \coquelicot{}
library, of the Lebesgue integral of non-negative
functions~\cite{boldo2021jar}.
This development is driven by detailed pencil-and-paper proofs written
for the purpose of formalization~\cite{clement2021arxiv}.
The theory of Lebesgue integration has been limited to non-negative
functions and stops at Tonelli's theorem~\cite{boldo2023fm} but it has
recently been extended with a formalization of the Bochner
integral~\cite{boldo2022coq}.
The authors have communicated to us that there is work in progress on
the Lebesgue measure but that it is not a modular construction like
ours.

The difference between the work by Boldo et al.\ and our work lies
more in the sustaining infrastructure than in the gallery of theorems.
First, we cannot reuse their framework because of many diverging
choices of conventions, one of them is assuming that
$\infty - \infty = 0$, which results in the addition of the extended
real numbers being non-associative, which prevents the use of iterated
operators {\it \`a la} \mathcomp{}~\cite[\sect{3.2}]{boldo2021jar}.
We also insist on developing abstractions and components developed
along \analysis{} so as to find the best encodings. For example, Boldo
et al.\ use a very concrete encoding of simple functions whose ranges
are represented by sorted
lists~\cite[\sect{6.3}]{boldo2021jar}. Notwithstanding the fact that
sorting is not essential to develop integration theory, it appears
that this makes for longer proofs. For example, we already
discussed the benefits of the infrastructure of iterated operators
over finite supports (Sect.~\ref{sec:fsbigop}) regarding the proof
that the integral of the sum of simple functions is the sum of the
integrals (Sect.~\ref{sec:sintegral}). The approach by Boldo et al.\
seems to make for a five times longer script (118 vs.\ 23 lines of codes,
see \L+LInt_SFp_plus+\footnote{\url{https://depot.lipn.univ-paris13.fr/mayero/coq-num-analysis/-/blob/d76dc70b06f70e2f1e99fd2ba3b22bba6ea78c91/Lebesgue/simple_fun.v\#L809}} \cite[file \L+simple_fun.v+]{milc}).
Another example having a \L+sigma_algebra+ predicate or a
\L+measurableType+ structure while Boldo et al.\ use the fact that a
class of sets is a \salgebra{} if and only if it is equal to the
smallest \salgebra{} generated by its elements.  We found this
characterization impractical in the presence of the hierarchy of
classes of sets, for which we need inheritance to work in order to
share theorem across structures.  With an inductive characterization,
theorems defined on a larger class of sets (e.g., semiring of sets)
could not be applied to a \salgebra{}.
On a related note, our definition of generated \salgebra{} in
Sect.~\ref{sec:generated_salgebra} generalizes the one by Boldo et
al.\ by defining the complement with respect to an arbitrary set
instead of the full set. This is very useful in practice to develop
the theory of measurable partial functions and in fine define the
Lebesgue integral as parameterized by a domain
(Sect.~\ref{sec:integral_measurable_function}).

The C-Corn library also deals with the formalization of integration in
\coq{} as it has a formalization of the fundamental theorem of
calculus~\cite{cruzfilipe2002types} but this is about the Riemann
integral and it is in a constructive setting.

The coq-proba library~\cite{coqproba} provides a formalization of the
Lebesgue measure and integral but limited to real-valued functions and
closed intervals.

\lean{} has an extensive formalization of measure and integration
theory. The main source of documentation is the code of
\mathlib{}~\cite{mathlib}.  To our understanding, measures are defined
as a special case of outer measures~\cite{vanDoorn2021itp}, following
the idea than any non-negative function can generate an outer measure
which in turn can generate the \salgebra{} of its Carath\'eodory
measurable sets.  Hence \mathlib{} does not have a hierarchy of
classes of sets reflecting the literature, as we did
(Sect.~\ref{sec:math_struct_measure}), even though we believe that they
naturally occur inside the proofs.
\mathlib{} provides the Lebesgue integral and its standard lemmas up
to Fubini's theorem and the Radon-Nikod{\'y}m theorem (which we have
also recently proved using our framework~\cite{ishiguro2023ppl}), and
is actually further generalized to the Bochner integral. It has also
supported the formalization of the Haar
measure~\cite{vanDoorn2021itp}, which generalizes the Lebesgue
measure.

We are not aware of a formalization of the Lebesgue measure or
integral in \nuprl~\cite{nuprl} or \agda~\cite{agda} which are
also proof assistants based on dependent type theory.

\paragraph*{About Measure and Integration Theory in Proof Assistants of the HOL Family}

The HOL family of proof assistants has several formalizations of
measure and integration theory. It can be traced back to a
formalization of measure theory in HOL in
2002~\cite[\sect{2.2.2}]{hurd2002phd} (work actually inspired by
earlier work in Mizar~\cite{bialas1992jfm}).  It was generalized in
HOL~4 2010~\cite[\sect{2.3}]{coble2010phd} and used to formalize
Lebesgue integration~\cite{coble2010phd,mhamdi2010itp}.  Work in HOL4
triggered a port in Isabelle/HOL that was eventually reworked in
2011~\cite[\sect{4.2}]{holzl2011itp}.
The Lebesgue measure is defined in Isabelle/HOL using the gauge
integral that was already available in Isabelle/HOL, i.e., it is not
built as an extension of a premeasure~\cite[\sect{4.6}]{holzl2011itp}.
This approach results from a port from
\hollight~\cite{harrison2013jar}.

\paragraph*{Measure and Integration Theory in Other Proof Assistants}
Proof assistants we have discussed so far are based on the LCF
approach which consists in having a small kernel to ensure the
soundness of proof checking. Other proof assistants based on an
augmented trusted base providing more automation have also been used
to formalize measure and integration theory.

The Mizar Mathematical Library (MML)~\cite{bancerek2018jar} provides a
formalization of measure theory that can be traced back to
1992~\cite{bialas1992jfm}.  The Lebesgue measure in MML has recently
been reconstructed~\cite{endou2020jfm} using an approach by extension
from a semialgebra of intervals to fix an earlier
formalization~\cite{bialas1995jfm}. This is of course in a very
different setting compared to our work in \coq{} since the Mizar proof
assistant relies on the Tarski–Grothendieck set theory instead of
dependent type theory.

NASALib~\cite{pvslib} also provides a construction of the Lebesgue
measure by extension but where extension is carried out from an
algebra of sets~\cite[file \L!hahn_kolmogorov.pvs!]{pvslib} instead of a semiring of sets as we do. NASALib
is written in PVS~\cite{owre1992cade}, an interactive and automated
prover based on higher-order logic that provides predicate subtyping
and dependent types~\cite{rushby1998ieee}.  The formalization of
measure theory and Lebesgue integration has been initiated in
2007~\cite{lester2007afm}.

To the best of our understanding, in \metamath{}~\cite{metamath}, the
Lebesgue measure is not defined by
extension\footnote{\url{https://us.metamath.org/mpeuni/df-vol.html}}.

We did not find a formalization of the Lebesgue measure or integration
in other mainstream theorem provers such as ACL2 or ProofPower, which
seems to be confirmed by the ``Formalizing 100 Theorems''
list~\cite{freek100}.

\section{Conclusion}
\label{sec:conclusion}

This paper introduced a \coq{} formalization of measure theory and
Lebesgue integration that is compatible with \mathcomp{} and that
extends \analysis{}.
This includes an original formalization of mathematical structures for
measure theory (Sect.~\ref{sec:math_struct_measure}), an original
formalization of the construction of measures using the Measure Extension
theorem (Sect.~\ref{sec:hahn_extension}), whose application to a
measure over a semiring of intervals yields the Lebesgue measure
(Sect.~\ref{sec:lebesgue_measure}).
This also allows for the construction of the Lebesgue integral and the
formalization of its theory up to Fubini's theorem
(Sect.~\ref{sec:lebesgue_integral}).

We argued about technical aspects of this formalization that we
believe improve on related work (Sect.~\ref{sec:related_work}).
At the beginning of this experiment, much work was dedicated to the
formalization of structures for measure theory and to enrich the
foundations (in particular, extended real numbers).
Our development now provides new reusable libraries of general
interest, in particular for extended real numbers and their sequences
(Sect.~\ref{sec:ereal}), sums over finite supports (Sect.~\ref{sec:fsbigop})
and over general sets (Sect.~\ref{sec:esum}).
As concrete applications that illustrate the reusability of our
formalization, we can mention the Lebesgue-Stieltjes measure, which
could be formalized using the same approach we used for the Lebesgue
measure in
Sect.~\ref{sec:lebesgue_measure}~\cite{lebesguestieltjesPR}, more
standard results about measure theory such as the Radon-Nikod{\'ym}
theorem~\cite{radonnikodymPR}, and the formalization of the semantics
of a probabilistic programming language~\cite{affeldt2023cpp}.

\paragraph*{Current and Future Work}
The \coq{} community now has several formalizations of integration,
that rely on different grounds.  We have been exchanging with the
members of the MILC project~\cite{milc} to look for ways to share the
development effort.
As a next step of our formalization, we plan to formalize the
fundamental theorem of calculus for the Lebesgue integral to connect
with the theory of derivatives of \analysis{}. We have also started
formalizing probability theory and in particular discrete random
variables to generalize existing work on the formalization of
discrete probabilities on top of \mathcomp{} (e.g.,
\cite{affeldt2020cs}) and to apply it to the formalization of
equational reasoning for probabilistic programming languages (e.g., to
extend~\cite{affeldt2021jfp} to continuous probabilities).


\bibliographystyle{abbrv}


\begin{thebibliography}{10}

\bibitem{abate2021csf}
C.~Abate, P.~G. Haselwarter, E.~Rivas, A.~V. Muylder, T.~Winterhalter,
  C.~Hritcu, K.~Maillard, and B.~Spitters.
\newblock {SSProve}: {A} foundational framework for modular cryptographic
  proofs in {Coq}.
\newblock In {\em 34th {IEEE} Computer Security Foundations Symposium ({CSF}
  2021), Dubrovnik, Croatia, June 21--25, 2021}, pages 1--15. {IEEE}, 2021.

\bibitem{affeldt2020ijcar}
R.~Affeldt, C.~Cohen, M.~Kerjean, A.~Mahboubi, D.~Rouhling, and K.~Sakaguchi.
\newblock Competing inheritance paths in dependent type theory: a case study in
  functional analysis.
\newblock In {\em In 10th International Joint Conference on Automated Reasoning
  (IJCAR 2020), Paris, France, June 29--July 6}, volume 12167(2) of {\em
  Lecture Notes in Artifical Intelligence}, pages 3--20. Springer, Jul 2020.

\bibitem{cohen2018jfr}
R.~Affeldt, C.~Cohen, and D.~Rouhling.
\newblock Formalization techniques for asymptotic reasoning in classical
  analysis.
\newblock {\em J.\ Formaliz.\ Reason.}, 11(1):43--76, 2018.

\bibitem{affeldt2023cpp}
R.~Affeldt, C.~Cohen, and A.~Saito.
\newblock Semantics of probabilistic programs using s-finite kernels in coq.
\newblock In {\em 12th {ACM} {SIGPLAN} International Conference on Certified
  Programs and Proofs ({CPP} 2023) Boston, MA, USA, January 16--17, 2023},
  pages 3--16. {ACM}, 2023.

\bibitem{affeldt2021jfp}
R.~Affeldt, J.~Garrigue, D.~Nowak, and T.~Saikawa.
\newblock A trustful monad for axiomatic reasoning with probability and
  nondeterminism.
\newblock {\em J.\ Funct.\ Program.}, 31(E17), 2021.

\bibitem{affeldt2020cs}
R.~Affeldt, J.~Garrigue, and T.~Saikawa.
\newblock Reasoning with conditional probabilities and joint distributions in
  {Coq}.
\newblock {\em Comput.\ Softw.}, 37(3):79--95, 2020.

\bibitem{lebesguestieltjesPR}
R.~Affeldt and Y.~Ishiguro.
\newblock Formalization of the {Lebesgue}-{Stieltjes} measure in
  {MathComp-Analysis}.
\newblock \url{https://github.com/math-comp/analysis/pull/677}, 2023.
\newblock Pull request to \cite{analysis}. Completed in 2022.

\bibitem{radonnikodymPR}
R.~Affeldt and Y.~Ishiguro.
\newblock Formalization of the {Radon}-{Nikod\'ym} theorem in
  {MathComp-Analysis}.
\newblock \url{https://github.com/math-comp/analysis/pull/818}, 2023.
\newblock Pull request to \cite{analysis}. Completed in 2022.

\bibitem{bancerek2018jar}
G.~Bancerek, C.~Bylinski, A.~Grabowski, A.~Kornilowicz, R.~Matuszewski,
  A.~Naumowicz, and K.~Pak.
\newblock The role of the {Mizar Mathematical Library} for interactive proof
  development in {Mizar}.
\newblock {\em J. Autom. Reason.}, 61(1-4):9--32, 2018.

\bibitem{bertot2008tphols}
Y.~Bertot, G.~Gonthier, S.~O. Biha, and I.~Pasca.
\newblock Canonical big operators.
\newblock In {\em 21st International Conference on Theorem Proving in Higher
  Order Logics (TPHOLs 2008), Montreal, Canada, August 18--21, 2008}, volume
  5170 of {\em Lecture Notes in Computer Science}, pages 86--101. Springer,
  2008.

\bibitem{bialas1992jfm}
J.~Bialas.
\newblock Properties of {Caratheodor}'s measure.
\newblock Technical report, 1992.
\newblock Formalized Mathematics 4.

\bibitem{bialas1995jfm}
J.~Bialas.
\newblock The one-dimensional {Lebesgue} measure.
\newblock Technical report, 1995.
\newblock Formalized Mathematics 7.

\bibitem{boldo2023fm}
S.~Boldo, F.~Cl{\'{e}}ment, V.~Martin, M.~Mayero, and H.~Mouhcine.
\newblock A {Coq} formalization of {Lebesgue} induction principle and
  {Tonelli}'s theorem.
\newblock In {\em 25th International Symposium on Formal Methods ({FM} 2023),
  L{\"{u}}beck, Germany, March 6--10, 2023}, volume 14000 of {\em Lecture Notes
  in Computer Science}, pages 39--55. Springer, 2023.

\bibitem{boldo2021jar}
S.~Boldo, F.~Clément, F.~Faissole, V.~Martin, and M.~Mayero.
\newblock A {Coq} formalization of {Lebesgue Integration} of nonnegative
  functions.
\newblock {\em J.\ Autom.\ Reason.}, 66(2):175--213, 2021.

\bibitem{boldo2022coq}
S.~Boldo, F.~Clément, and L.~Leclerc.
\newblock A {Coq} formalization of the {Bochner} integral, 2022.
\newblock arXiv cs.LO 2201.03242.

\bibitem{clement2021arxiv}
F.~Clément and V.~Martin.
\newblock {Lebesgue} integration, detailed proofs to be formalized in {Coq},
  2021.
\newblock arXiv cs.LO 2101.05678.

\bibitem{coble2010phd}
A.~R. Coble.
\newblock {\em Anonymity, information, and machine-assisted proof}.
\newblock PhD thesis, University of Cambridge, King’s College, Jul 2010.
\newblock TR UCAM-CL-TR-785.

\bibitem{cohen2012phd}
C.~Cohen.
\newblock {\em Formalized algebraic numbers: construction and first-order
  theory}.
\newblock PhD thesis, École Doctorale de l’École Polytechnique, Laboratoire
  d’Informatique de l’École Polytechnique, Nov 2012.

\bibitem{cohen2020fscd}
C.~Cohen, K.~Sakaguchi, and E.~Tassi.
\newblock {Hierarchy} {Builder}: Algebraic hierarchies made easy in {Coq} with
  {Elpi} (system description).
\newblock In {\em 5th International Conference on Formal Structures for
  Computation and Deduction ({FSCD} 2020), June 29--July 6, 2020, Paris, France
  (Virtual Conference)}, volume 167 of {\em LIPIcs}, pages 34:1--34:21. Schloss
  Dagstuhl - Leibniz-Zentrum f{\"{u}}r Informatik, 2020.

\bibitem{nuprl}
R.~L. Constable, S.~F. Allen, M.~Bromley, R.~Cleaveland, J.~F. Cremer,
  R.~Harper, D.~J. Howe, T.~B. Knoblock, N.~P. Mendler, P.~Panangaden, J.~T.
  Sasaki, and S.~F. Smith.
\newblock {\em Implementing mathematics with the Nuprl proof development
  system}.
\newblock Prentice Hall, 1986.

\bibitem{cruzfilipe2002types}
L.~Cruz{-}Filipe.
\newblock A constructive formalization of the fundamental theorem of calculus.
\newblock In {\em Selected Papers of the Second International Workshop on Types
  for Proofs and Programs ({TYPES} 2002), Berg en Dal, The Netherlands, April
  24--28, 2002}, volume 2646 of {\em Lecture Notes in Computer Science}, pages
  108--126. Springer, 2002.

\bibitem{endou2020jfm}
N.~Endou.
\newblock Reconstruction of the one-dimensional {Lebesgue} measure.
\newblock Technical report, National Institute of Technology, Gifu College,
  2020.
\newblock Formalized Mathematics 28(1):93--104.

\bibitem{garillot2009tphols}
F.~Garillot, G.~Gonthier, A.~Mahboubi, and L.~Rideau.
\newblock Packaging mathematical structures.
\newblock In {\em 22nd International Conference on Theorem Proving in Higher
  Order Logics (TPHOLs 2009), Munich, Germany, August 17--20, 2009}, volume
  5674 of {\em Lecture Notes in Computer Science}, pages 327--342. Springer,
  2009.

\bibitem{gonthier2013itp}
G.~Gonthier, A.~Asperti, J.~Avigad, Y.~Bertot, C.~Cohen, F.~Garillot, S.~L.
  Roux, A.~Mahboubi, R.~O'Connor, S.~O. Biha, I.~Pasca, L.~Rideau, A.~Solovyev,
  E.~Tassi, and L.~Th{\'{e}}ry.
\newblock A machine-checked proof of the odd order theorem.
\newblock In {\em 4th International Conference on Interactive Theorem Proving
  ({ITP} 2013), Rennes, France, July 22--26, 2013}, volume 7998 of {\em Lecture
  Notes in Computer Science}, pages 163--179. Springer, 2013.

\bibitem{harrison2013jar}
J.~Harrison.
\newblock The {HOL} light theory of euclidean space.
\newblock {\em J. Autom. Reason.}, 50(2):173--190, 2013.

\bibitem{holzl2011itp}
J.~H{\"{o}}lzl and A.~Heller.
\newblock Three chapters of measure theory in {Isabelle}/{HOL}.
\newblock In {\em Second International Conference on Interactive Theorem
  Proving ({ITP} 2011), Berg en Dal, The Netherlands, August 22--25, 2011},
  volume 6898 of {\em Lecture Notes in Computer Science}, pages 135--151.
  Springer, 2011.

\bibitem{hurd2002phd}
J.~Hurd.
\newblock {\em Formal Verification of Probabilistic Algorithms}.
\newblock PhD thesis, University of Cambridge, 2002.
\newblock UCAM-CL-TR-566.

\bibitem{ishiguro2023ppl}
Y.~Ishiguro and R.~Affeldt.
\newblock A progress report on formalization of measure theory with
  {MathComp-Analysis}.
\newblock In {\em 25th Workshop on Programming and Programming Languages
  (PPL2023), Nagoya University, March 6--8, 2023}. Japan Society for Software
  Science and Technology, Mar 2023.

\bibitem{klenke2014}
A.~Klenke.
\newblock {\em Probability Theory: A Comprehensive Course}.
\newblock Springer, 2014.
\newblock 2nd edition.

\bibitem{milc}
{Le projet MILC}.
\newblock Numerical analysis in {Coq}.
\newblock \url{https://depot.lipn.univ-paris13.fr/mayero/coq-num-analysis},
  2023.
\newblock Since 2018. See also \url{https://lipn.univ-paris13.fr/MILC}.

\bibitem{lester2007afm}
D.~R. Lester.
\newblock Topology in {PVS}: Continuous mathematics with applications.
\newblock In {\em 2nd Workshop on Automated Formal Methods (AFM 2007)}, pages
  11--20. Association for Computing Machinery, 2007.

\bibitem{liintegration}
D.~Li.
\newblock {\em Intégration et applications---Cours et exercices corrigés}.
\newblock Eyrolles, 2016.

\bibitem{mathcomp}
{Mathematical Components Team}.
\newblock {Mathematical} {Components} library.
\newblock \url{https://github.com/math-comp/math-comp}, 2007.
\newblock Last stable version: 2.0 (2023).

\bibitem{metamath}
N.~Megill.
\newblock {\em Metamath: A Computer Language for Mathematical Proofs}.
\newblock 2019.
\newblock Available at \url{https://us.metamath.org/downloads/metamath.pdf}.
  With extensive revisions by David A.\ Wheeler.

\bibitem{mhamdi2010itp}
T.~Mhamdi, O.~Hasan, and S.~Tahar.
\newblock On the formalization of the {Lebesgue} integration theory in {HOL}.
\newblock In {\em First International Conference on Interactive Theorem Proving
  ({ITP} 2010), Edinburgh, UK, July 11--14, 2010}, volume 6172 of {\em Lecture
  Notes in Computer Science}, pages 387--402. Springer, 2010.

\bibitem{owre1992cade}
S.~Owre, J.~M. Rushby, and N.~Shankar.
\newblock {PVS:} {A} prototype verification system.
\newblock In {\em 11th International Conference on Automated Deduction
  (CADE-11), Saratoga Springs, NY, USA, June 15--18, 1992}, volume 607 of {\em
  Lecture Notes in Computer Science}, pages 748--752. Springer, 1992.

\bibitem{rushby1998ieee}
J.~M. Rushby, S.~Owre, and N.~Shankar.
\newblock Subtypes for specifications: Predicate subtyping in {PVS}.
\newblock {\em {IEEE} Trans. Software Eng.}, 24(9):709--720, 1998.

\bibitem{coqproba}
J.~Tassarotti, J.~Tristan, and K.~Palmskog.
\newblock coq-proba: A probability theory library for the {Coq} theorem prover.
\newblock \url{https://github.com/jtassarotti/coq-proba}, 2023.
\newblock Since 2019.

\bibitem{tassarotti2021cpp}
J.~Tassarotti, K.~Vajjha, A.~Banerjee, and J.~Tristan.
\newblock A formal proof of {PAC} learnability for decision stumps.
\newblock In {\em 10th {ACM} {SIGPLAN} International Conference on Certified
  Programs and Proofs (CPP 2021), Virtual Event, Denmark, January 17--19,
  2021}, pages 5--17. {ACM}, 2021.

\bibitem{agda}
{The Agda Team}.
\newblock {\em Agda's documentation v2.6.3}, 2023.
\newblock Available at \url{https://agda.readthedocs.io/en/v2.6.3}.

\bibitem{coq}
{The Coq Development Team}.
\newblock {\em The {Coq} Proof Assistant Reference Manual}.
\newblock Inria, 2023.
\newblock Available at \url{https://coq.inria.fr/refman/}. Version 8.17.0.

\bibitem{analysis}
{The MathComp-Analysis Team}.
\newblock {MathComp-Analysis}: Mathematical components compliant analysis
  library.
\newblock \url{https://github.com/math-comp/analysis}, 2023.
\newblock Since 2017. Last stable version: 0.6.2. This paper refers to the
  branch {\tt hierarchy-builder}.

\bibitem{mathlib}
{The mathlib community}.
\newblock {Lean} mathematical components library.
\newblock \url{https://github.com/leanprover-community/mathlib}, 2023.
\newblock Since 2017.

\bibitem{pvslib}
{The NASALib development team}.
\newblock {NASA} {PVS} library of formal developments.
\newblock Current version: 7.1.1. Available at
  \url{https://github.com/nasa/pvslib}., 2023.

\bibitem{vanDoorn2021itp}
F.~van Doorn.
\newblock Formalized {Haar} measure.
\newblock In {\em 12th International Conference on Interactive Theorem Proving
  ({ITP} 2021) June 29--July 1, 2021, Rome, Italy (Virtual Conference)}, volume
  193 of {\em LIPIcs}, pages 18:1--18:17. Schloss Dagstuhl - Leibniz-Zentrum
  f{\"{u}}r Informatik, 2021.

\bibitem{freek100}
F.~Wiedijk.
\newblock Formalizing 100 theorems.
\newblock \url{http://www.cs.ru.nl/~freek/100}, 2023.

\end{thebibliography}

\end{document}